\documentclass{elsart}%
\usepackage{amssymb}
\usepackage{amsfonts}
\usepackage{amsmath}
\usepackage{graphicx}%
\begin{document}
%
\begin{frontmatter}%
%

\title
{Generalized Fractal Kinetics in Complex Systems  ( Application to Biophysics and Biotechnology)}%
%

\author{F.Brouers and O.Sotolongo-Costa}%
%

\address{Department of Physics, University of Liège, 4000, Belgium}%
%

\address
{Faculty of Physics, Chair of Complex Systems H.Poincaré, University of Havana, Cuba}%
%

\address{fbrouers@ulg.ac.be,    oscarso@fisica.uh.cu}%
%

\begin{abstract}%
We derive a universal function for the kinetics of complex systems
characterized by stretched exponential and/or power-law behaviors.This kinetic
function unifies and generalizes previous theoretical attempts to describe
what has been called "fractal kinetic".

The concentration evolutionary equation is formally similar to the relaxation
function obtained in the stochastic theory of relaxation, with two exponents
$\alpha$ and $n$.\ The first one is due to memory effects and short-range
correlations and the second one finds its origin in the long-range
correlations and geometrical frustrations which give rise to ageing behavior.
These effects can be formally handled by introducing adequate probability
distributions for the rate coefficient.\ We show that the distribution of rate
coefficients is the consequence of local variations of the free energy (energy
landscape) appearing in the exponent of the Arrhenius formula.

The fractal $(n,\alpha)$ kinetic is the applied to a few problems of
fundamental and practical importance in particular the sorption of dissolved
contaminants in liquid phase. Contrary to the usual practice in that field, we
found that the exponent $\alpha$, which is implicitly equal to $1$ in the
traditional analysis of kinetic data in terms of first or -second order
reactions, is a relevant and useful parameter to characterize the kinetics of
complex systems. It is formally related to the system energy landscape which
depends on physical, chemical and biological internal and external factors.

We discuss briefly the relation of the $(n,\alpha)$ kinetic formalism with the
Tsallis theory of nonextensive systems.
\end{abstract}%
%

%

\end{frontmatter}%

\bigskip

Introduction

The physical origin of anomalous kinetics in complex systems like glass,
polymers, solutions, proteins, enzymes... has received much attention in
recent years (\cite{Kop88}\cite{Sav95}\cite{Fra99}\cite{Wer99}\cite{Sch04}).

This is due to the observation that in many instances, the kinetics cannot be
accounted for without introducing a time-dependent reaction rate coefficient
to describe properly the experimental data. As in many other related problems
an accurate relation between microscopic properties and the global macroscopic
observables is lacking due to many-body interactions, the inevitable
homogenization and coarse-graining resulting from complexity and experimental
techniques. What has been done in practice to deal with this situation is to
generalize formula used in simple reaction kinetics (first and second order
reaction) and introduce one or several supplementary empirical parameters to
fit experimental data.\ The aim of these works is to establish correlations
between macroscopic observables and external parameters (temperature,
concentration, pH, ...). Another method widely used is to consider a few
reaction steps.\ The total rate equation is written as the sum of elementary
rate equations (first or second order).\ Their respective weight and the rate
coefficients are fitted to the data. This procedure can lead to ambiguous
results and conclusions. In particular it does not reproduce power-law
behaviors often encountered in these systems. In this spirit, Frauenfelder and
collaborators \cite{Fra99} have used intuitive arguments to fit experimental
biomolecular reaction data in protein materials with empirical stretched
exponential or power-law functions.\ They trace the 'anomalous" kinetics to
the distribution of the reaction rate in the Arrehnius formula. In 1988, to
account for power law behavior, Kopelman \cite{Kop88} proposed a
phenomenological fractal like kinetics to account for reactions in materials
prepared as fractals. This lead more recently Savageau \cite{Sav95} to
introduce a model where instead of introducing a time-dependence to the rate
coefficient, the reactant concentrations are raised to non-integer powers.

More recently, Weron \cite{Wer99} and collaborators using results from their
stochastic theory of relaxation in dipolar systems \cite{Wer97}, introduced a
time dependent power-law reaction rate coefficient to generalize first and
second order kinetic equations in order to apply them to biomolecular
reactions. Simultaneously, Mendes and collaborators \cite{Men99}, using
results of the Tsallis nonextensive entropy theory \cite{Tsa01} to solve
non-linear differential equations, introduced the concept of a $n$-order
kinetic equation whose solution has a formal expression similar to the Tsallis
generalized Pareto distribution.

The purpose of the present paper is to use some results of two recent papers
on non-Debye relaxation \cite{Bro04}\cite{Bro05b} to incorporate the ideas
developed in the previous quoted works \cite{Kop88}\cite{Fra99}\cite{Wer99}%
\cite{Men99}, in one unified formalism in order to the set the basis of a
general theory of reaction kinetics in complex systems.

The challenge which has to be faced, in this important problem, is to give a
physical or statistical meaning to the empirical parameters and, when this is
possible, to relate the "anomalous" kinetics to universality, i.e. universal
scaling functions independent on the microscopic or mesoscopic detailed
properties of the system.

\section{The ($n,\alpha)$ kinetic equation}

The most general chemical kinetic equation for one given species ($A$) in a
complex system composed of $A,B,C,$..\ reacting atoms and molecules can be
written formally as
\begin{equation}
\frac{dc_{A}}{dt}=Kc_{A}^{\alpha}c_{B}^{\beta}c_{C}^{\gamma}...
\end{equation}
where $K$ is the rate coefficient and $\alpha,\beta,\gamma$...refer to the
concentrations of chemical species $A,B,C,...$present in the reaction and the
sum $\alpha+\beta+\gamma$ is the overall order of the reaction.\ In some
cases, the concentration $c_{B,}c_{C}...$ can be considered as constant, thus
the above equation reduces (for instance for the reactant $A)$ to the form
\begin{equation}
-\frac{dc_{n}(t)}{dt}=K_{n}c_{n}(t)_{\ }^{n}%
\end{equation}
In that way, the parameter $n$ becomes the overall order of the reaction. The
solution of this differential equation for one of the reactants is given by
\cite{Men99}%
\begin{equation}
c_{n}(t)=c_{n}(0)[(1+(n-1)c_{n}(0)^{n-1}K_{n}t]^{-\frac{1}{n-1}}%
\end{equation}
which has the form of a generalized Pareto function, solution of the Tsallis
entropy maximization \cite{Tsa01} and has an asymptotic power law behavior
$c_{n}(t)\propto t^{-1/(n-1)}$.

If we use the deformed $n$-exponential and $n$-logarithm introduced by Tsallis
and collaborators in the context of nonextensive systems \cite{Tsa01}:%
\[
\exp_{n}(x)=(1-(n-1)x)^{-\frac{1}{n-1}}\text{ \ \ if \ \ }1-(n-1)x\text{\ }%
>0,\ \ 0\text{ otherwise}%
\]%
\begin{equation}
\text{and\ \ \ \ ln}_{n}(x)=\frac{x^{1-n}-1}{1-n}\
\end{equation}
with%
\[
\exp_{n}(\text{\ ln}_{n}(x))\text{\ =}\ln_{n}(\text{\ exp}_{n}(x))=1
\]
we can write the solution (3) in a more compact form:%
\begin{equation}
c_{n}(t)=c_{n}(0)\exp_{n}(-t/\tau_{n})
\end{equation}
with a characteristic time:%
\begin{equation}
\tau_{n}=(c_{n}(0)^{n-1}K_{n})^{-1}%
\end{equation}
This will allow the definition of an effective time-dependent rate
coefficient:%
\begin{align}
-\frac{dc_{n}(t)}{dt}  &  =c(0)\frac{1}{\tau_{n}}(\exp_{n}(-(t/\tau_{n}%
))^{n}\\
&  =c(0)\frac{1}{\tau_{n}}(1+(n-1)(t/\tau_{n}))^{-\frac{1}{n-1}-1}\nonumber\\
&  =\mathcal{K}_{n}(t)c_{n}(t)\nonumber
\end{align}
with%
\begin{equation}
\mathcal{K}_{n}(t)=\frac{1}{\tau_{n}}(1+(n-1)(t/\tau_{n}))^{-1}%
\end{equation}
For $t<<\tau_{n},$ one has a slowing down of the effective rate :%
\begin{equation}
\mathcal{K}_{n}(t)=\frac{1}{\tau_{n}}(1-(n-1)(t/\tau_{n}))\text{
\ \ \ \ }t<<\tau_{n}\text{\ \ }%
\end{equation}
and for $n\neq1$, $t>>\tau_{n}$%
\begin{equation}
\mathcal{K}_{n}(t)\propto(1/(n-1))t^{-1}%
\end{equation}
This behavior is a manifestation of what has been call aging \cite{Par97}
which appears as soon as $n\neq1$. For $n=1$, one recovers the exponential
behavior with $\mathcal{K}(t)=\ 1/\tau.$

These results do not exhibit the $t<\tau$ power law time dependence of the
reaction rate which describes adequately the experimental data of many complex
systems \cite{Wer99}\cite{Plo86}\cite{Fra94}\cite{Ber96}\cite{Dew97}.\ This
behavior can appear quite naturally if we introduce in (6), instead of the
$n$-exponential, a $n-$Weibull function introduced by Mendes \cite{Men04} and
used also in the theory of relaxation \cite{Bro04}\cite{Bro05b}
\begin{equation}
c_{n,\alpha}(t)=c(0)\exp_{n}(-(t/\tau_{n,\alpha})^{\alpha}%
)=c(0)[(1+(n-1)\ (t/\tau_{n,\alpha})^{\alpha}]^{-\frac{1}{n-1}}%
\end{equation}
with a characteristic time:%
\begin{equation}
\tau_{n,\alpha}=[K_{n,\alpha}c(0)^{n-1}]^{-1/_{\alpha}}%
\end{equation}
The effective time-dependent rate coefficient $\mathcal{K}_{n,\alpha}(t)$ now
reads
\begin{equation}
\mathcal{K}_{n,\alpha}(t)=\alpha\frac{t^{\alpha-1}}{\tau_{n,\alpha}^{\alpha}%
}(1+(n-1)(t/\tau_{n,\alpha})^{\alpha})^{-1}%
\end{equation}
Equation (11) is solution of a fractional differential equation :%
\begin{equation}
-\frac{dc_{n,\alpha}(t)}{dt^{\alpha}}=K_{n,\alpha}c_{n,\alpha}(t)_{\ }^{n}%
\end{equation}
by introducing a fractional time index $\alpha$ and a non-integer reaction
order $n$. Fractional derivation and fractal time concepts have been
introduced in physics (diffusion in disordered and porous media, random walks
...\cite{Hil200}) in the theory of dielectric response \cite{Nik87}%
\cite{Sch88}\cite{Bro05b} and in economy \cite{Man97}. \ 

The effective rate coefficient $\mathcal{K}_{n,\alpha}(t)$ has the two
asymptotic behaviors%
\begin{align}
\text{for \ \ }t  &  \rightarrow0\text{ \ \ \ }\mathcal{K}_{n,\alpha
}(t)\propto t^{\alpha-1}\text{\ \ }\\
\text{for \ \ }t  &  \rightarrow\infty\text{ \ \ \ }\mathcal{K}_{n,\alpha
}(t)\propto t^{-1}\text{\ }\nonumber
\end{align}
For \ $t\rightarrow0,$ we get the same power-law variation of the rate
coefficient as in the work of Weron \textit{et al}. \cite{Wer99} as well as in
the fractal phenomenological description of non-homogeneous reaction dynamics
called fractal-like kinetics \cite{Kop88}, if we identify the Kopelman fractal
parameter $h<1$ with $1-\alpha.$ As noted in \cite{Sch04}, the concept of
effective time-dependent rate constant breaks down for $t\rightarrow0,$
$\alpha<1$, since in that limit $\mathcal{K}_{n,\alpha}(t) $ diverges. The
general solution (11) of the fractional differential equation (14)\ does not
suffer from such difficulty and is well defined in the positive time domain.
In any case, as for real geometric fractals, for physical reasons, there is in
each case a natural small time cut-off.

The two asymptotic behaviors of the concentration evolutionary law equation
(11) are:%
\begin{equation}
c_{n,\alpha}(t)=c(0)[(1-\ (t/\tau_{n,\alpha})^{\alpha}+...]
\end{equation}
independent of $n$ for $t<<\tau_{n,\alpha}$, while for $n\neq1$ and
$t>>\tau_{n,\alpha}$
\begin{equation}
c_{n,\alpha}(t)=c(0)(n-1)\ (t/\tau_{n,\alpha})^{-\alpha/(n-1)}%
\end{equation}
The ratio of the the two asymptotic exponents $\alpha$ and $\alpha/(n-1)$
yields the value of the apparent order of the reaction $n$.

For special values of the two parameters $n$ and $\alpha$, some other typical
solutions are recovered \ 

a. If $n=1,\alpha=1$, we have%
\begin{equation}
-\frac{dc(t)}{dt}=K_{1}c(t)\text{ \ }\rightarrow\text{ \ }c(t)=c(0)\exp
(-K_{1}t)
\end{equation}
which is a first order kinetic

b. If $n=1,\alpha\neq1$, we have%
\begin{equation}
-\frac{dc_{\alpha}(t)}{dt^{\alpha}}=K_{\alpha}c_{\alpha}(t)\rightarrow\text{
\ }c_{\alpha}(t)=c(0)\exp(-K_{\alpha}t^{\alpha})
\end{equation}
which is a "Weibull kinetics".\ If $\ 0<\alpha<1,$ it is a "stretched
exponential kinetic".

c.\ If $n\neq1$, $\alpha=1,$ equations (11)\ gives
\begin{equation}
c_{q}(t)=c_{n}(0)[(1+(n-1)c_{n}(0)^{n-1}K_{n}t]^{-\frac{1}{n-1}}\equiv\exp
_{n}(-c(0)^{n-1}K_{n}t)
\end{equation}
which is solution of (2).

\ d. If $n=2,$ $\alpha=1,$ we have%
\begin{equation}
-\frac{dc(t)}{dt}=K_{2}c^{2}(t)\text{ \ }\rightarrow\text{ \ }%
1/c(t)-1/c(0)=K_{2}t
\end{equation}
This is the second order kinetic.

e. If $n=2$, $\alpha\neq1,$ we have%
\begin{equation}
-\frac{dc_{\alpha}(t)}{dt^{\alpha}}=K_{2,\alpha}c_{\alpha}(t)^{2}\text{
\ \ }\rightarrow c_{\alpha}(t)=c(0)[(1+\ c(0)(K_{2,\alpha}t)^{\alpha}]^{-1}%
\end{equation}
This is a generalized second order kinetic.

Cases (b) and (e) have been discussed in \cite{Wer99}.

It is important to note that as soon as $n\neq1$, the time dependence of the
kinetics depends on the initial concentration.

We will call the kinetic giving rise to the concentration evolutionary law
(11), the $(n,\alpha)$ kinetic:
\begin{equation}
c_{n,\alpha}(t)=c(0)[(1+(n-1)\ (t/\tau_{n,\alpha})^{\alpha}]^{-\frac{1}{n-1}}%
\end{equation}%
\begin{equation}
\tau_{n,\alpha}=\ [c(0)^{n-1}K_{n,\alpha}]^{-1/\alpha}%
\end{equation}
is the characteristic time of the complex kinetic. It depends on the initial
concentration and the two exponents $n$ and $\alpha.\ $For $n\rightarrow1$,
$c_{n,\alpha}(t)$ tends to a Weibull exponential with $\tau_{1,\alpha
}=[K_{1,\alpha}]^{-1/\alpha}$ . One can define a "half-reaction time"
$\tau_{1/2}$ which is the time necessary to transform half of the relevant
quantity by solving the equation%
\begin{equation}
(1+(n-1)(\tau_{1/2}/\tau_{n,\alpha})^{\alpha})\ ^{-1/(n-1)}=1/2
\end{equation}
which gives using the definition of $\ln_{n}(x)$ (4) :%
\begin{equation}
\tau_{1/2}=\tau_{n,\alpha}(\ln_{n}2)^{1/_{\alpha}}%
\end{equation}
Kinetics are "memoryless" only when $n=\alpha=1.$ If $n=1,$ kinetics are
"memoryless" in the fractal time $t_{f}=t^{\alpha}$ since with the change of
variable, the Weibull function reduces to a memoryless exponential.

One can introduce in this problem a "response function" as it is done in Weron
et al.\cite{Wer99} for $n=1$ and $n=2.$ We have more generallyfor any real
$n$:%
\begin{equation}
f(t)=-\frac{1}{c(0)}\frac{dc_{n,\alpha}(t)}{dt}=\alpha\frac{t^{\alpha-1}}%
{\tau_{n,\alpha}}(1+(n-1)(\frac{t}{\tau_{n,\alpha}})^{\alpha})^{-\frac{1}%
{n-1}-1}%
\end{equation}
this function has the two asymptotic behaviors:%
\begin{align}
\text{for \ \ }t  &  \rightarrow0\text{ \ \ \ }f(t)\propto(t/\tau_{n,\alpha
})^{\alpha-1}\text{\ \ }\\
\text{for \ \ }t  &  \rightarrow\infty\text{ \ \ \ \ }f(t)\propto
(t/\tau_{n,\alpha})^{-(\alpha/n-1)-1}\text{\ \ }n\neq1\nonumber
\end{align}
For $n=1$ and $n=2$, they coincide with those of \cite{Wer99}.

\section{Arrhenius law and exponential conspiracy}

The results of the previous section can be understood physically as a
consequence of what is has been called "exponential conspiracy", an expression
coined by Boucheau \cite{Bou96} and proposed as exercise in textbooks on
probability theory(for example \cite{Foa98}). It is generally accepted that
the temperature dependence of the reaction rate $K$ has an Arrhenius form
which we will write:%
\begin{equation}
K=K_{0}\exp(\pm E/kT)
\end{equation}
$K_{0}$ is the pre-exponential factor and $E$ the relevant energy (in
thermodynamics systems this energy is the Gibbs free energy which depends on
the enthalpy (heat of reaction) and the entropy : $\ G=H-TS$).\ The sign +
corresponds to an "exothermic" reaction (i.e. the energy corresponds to an
attraction energy and the rate decreases with the temperature).\ This is the
case for instance in physisorption, when the overall adsorption enthalpy
resulting from adsorption and desorption is positive.\ The sign - corresponds
to an "endothermic" reaction and $E$ is an activation energy barrier to be
overcome. In that case the rate increases with temperature. We have written
the two terms "exothermic" and "endothermic", because due to the variation of
entropy with $T,$ paradoxically in some complex systems, an endothermic
reaction can occur without activation energy (see for instance \cite{Ho03}).

In disordered systems frozen out of equilibrium, the exponent factor $E/kT$
varies due to fluctuations\ of local energies and\ local temperatures. The
distribution of energies depends on what has been called by Frauenfelder
\cite{Fra99} the "energy landscape", a concept taken from the theory of
glasses, and variations of the inverse of local temperature $(1/T)$ have been
used to introduce, what has been called "super-statistics" by Beck and Cohen
\cite{Bec03}.\ 

Here we will assume, as in the theory of heterogeneous catalysis, in the
theory of glass and in the theory of adsorption \cite{Bro05a}, that the
probability distribution of the energy $E$ varies for large values as
$\exp(-E/E_{0}).\ $This means that the large energy values are statistically
exponentially small.\ The reference energy $E_{0}$ is linked to the width of
the energy density distribution $f_{E}(E)$ \cite{Bro05a}. With this
assumption, using the basic probability theory relation%
\begin{equation}
f_{E}(E)dE=f_{K}(K)dK
\end{equation}
it is straightforward to show, that the distribution of the rate coefficient
$K$, has the asymptotic form%
\begin{equation}
f_{K}(K)=\mu(K/K_{0})^{-1\pm\mu}\text{ \ \ \ with \ \ }\mu=kT/E_{0}%
\end{equation}
Therefore, in the "exothermic" case (sign + in (18)),
\begin{equation}
\text{ }K\rightarrow\infty\text{ \ \ \ }f_{K}(K)\sim K^{-1-\mu}%
\end{equation}
while in the "endothermic" case (sign - in (18))
\begin{equation}
\text{ }K\rightarrow0\text{ \ \ \ }f_{K}(K)\sim K^{-1+\mu}%
\end{equation}
In the first case, the distribution $f_{K}(K)$ is a Pareto distribution and is
the simplest density distribution belonging to the domain of attraction of the
stable L\'{e}vy distributions \cite{Sor04}\cite{Ucha99}. If we assume that the
variations of $K$ induced by the fluctuations of $E$ are represented by a
L\'{e}vy distribution $L_{\mu}(\lambda)$ , one can, using the well-known
relation (the Laplace transform of a one-sided Levy distribution is a
stretched exponential) :
\begin{equation}
\int_{0}^{\infty}\exp(-\lambda Kt)\mathcal{L}_{\mu}\ (\lambda)d\lambda
=\exp(-(Kt)^{\mu})
\end{equation}
obtain the generalized first order (Weibull) kinetic (case 2 with $\mu=\alpha
$) as a compounded exponential first order kinetic . In the second case, if we
use the Gamma density distribution which has the power law asymptotic behavior
($\mu\lambda^{\mu-1}$) for small values of $\lambda$,%
\begin{equation}
g_{\mu}(\lambda)=\frac{\mu}{\ \Gamma(\mu)}(\mu\lambda)^{\mu-1}\exp(-\mu
\lambda)
\end{equation}
we obtain equation (3) with $\mu=1/(n-1)$%
\begin{equation}
\int_{0}^{\infty}\exp(-\lambda(Kt))g_{\mu}(\lambda)d\lambda=(1+\frac{1}{\mu
}(Kt))^{-\mu}%
\end{equation}
If we use the Weibull distribution, we can then write \cite{Rod77}%
\begin{equation}
\int_{0}^{\infty}\exp(-\lambda(Kt)^{\alpha})g_{\mu}(\lambda)d\lambda
=(1+\frac{1}{\mu}(Kt)^{\alpha})^{-\mu}%
\end{equation}
or using a result established by Weron and collaborators in the stochastic
theory of relaxation \cite{Wer97} :
\begin{equation}
\int_{0}^{\infty}\exp(-\lambda Kt){\scriptsize \ }ML_{\alpha,\mu}%
(\lambda)d\lambda=(1+(1/\mu)(Kt)^{\alpha})^{-\mu}%
\end{equation}
where $ML_{\alpha,\mu}(\lambda)$ is a generalized Mittag-Lefller
distribution:
\begin{equation}
ML_{\alpha,\mu}(\lambda)=\sum_{k=0}^{\infty}\frac{(-1)^{k}\Gamma(\mu
+k)}{k!\Gamma(\mu)\Gamma\lbrack(\alpha(\mu+k)]}(\lambda)^{\alpha(\mu+k)-1}%
\end{equation}
This last result is more difficult to interpret. It can be understood
\cite{Wer97}\cite{Bro04}, as the result of the random character of the number
of active centers, geometric frustrations and dynamic constraints or as a
consequence of the interplay of "exothermic" and "endothermic" processes in
the kinetics of complex materials. In conclusion, the $(n,\alpha)$ kinetic
equation, can be obtained formally by introducing an adequate distribution for
the exponent of the Arrehnius law as conjectured by Fraunfelder \cite{Fra99}.

\section{Probabilistic interpretation of the $(n,\alpha)$ kinetic equation}

A comparison with the stochastic theory of relaxation \cite{Wer97}%
\cite{Wer99}\cite{Bro04}\cite{Bro05b} is of interest to understand the
physical meaning of equation (23). We first note that%
\begin{equation}
c_{n,\alpha}(t)=c(0)[(1+(n-1)\ (t/\tau_{n,\alpha})^{\alpha}]^{-\frac{1}{n-1}}
\tag{23}%
\end{equation}
is related to the Burr$_{XII}$ distribution function $(B_{a,b,c}%
(x)=1-(1+ax^{b})^{-c}$ $\ ,x>0)$ \cite{Rod77}, named by reference to the
number it occupies in the main Table of Burr's original paper \cite{Bur42}).
If we introduce an effective random reaction waiting time $\tilde{\theta}$
,\ the quantity $c_{n,\alpha}(t)/c(0)$ can be viewed as the probability that
the reactant has not yet reacted at time $t$:%
\begin{equation}
c_{n,\alpha}(t)/c(0)=\Pr(\tilde{\theta}>t)=1-\int_{0}^{t}f_{n,\alpha}%
(\tilde{\theta})d\tilde{\theta}%
\end{equation}
where%
\begin{equation}
f_{n,\alpha}(\tilde{\theta})=\alpha\frac{\tilde{\theta}^{\alpha-1}}%
{\tau_{n,\alpha}}(1+(n-1)(\frac{\tilde{\theta}}{\tau_{n,\alpha}})^{\alpha
})^{-\frac{1}{n-1}-1}%
\end{equation}
This distribution belongs to the domain of attraction of the Levy distribution
with a tail exponent $\mu=\alpha/(n-1)$ and therefore generalizes the Pareto
or Zipf-Mandelbrot distributions used in fractal reactions kinetics of
previous works \cite{Sch04}. If $\mu<1,$ an expectation value of
$\tilde{\theta}$ cannot be defined and an escort probability function
\cite{Tsa98} has to be used to determine $\tau_{n,\alpha}$ from the knowledge
of $f_{n,\alpha}(\tilde{\theta})$.

If $\alpha=1,$ the probability density function (41) reduces to the Tsallis
generalized Pareto distribution \cite{Tsa01} if ($n-1)=$ $(q-1)/(2-q)$ i.e.
$n=1/(2-q)$
\[
f_{q}(\tilde{\theta})=\frac{\ 1}{\tau_{n,\alpha}}(1+(q-1)/(2-q)\frac
{\tilde{\theta}}{\tau_{n,\alpha}})^{-\frac{1}{q-1}}%
\]
which maximizes the Tsallis entropy of the random variable $\tilde{\theta}.$

The relation between the $(n,\alpha)$ kinetics and the nonextensivity of the
entropy and the formal relation of the reaction order $n$ with the Tsallis
entropy index $q$ is worth further investigations. The characteristics of the
complex systems studied in the present work (see the introduction) are similar
to the ones (frozen non-equilibrium states with memory effects and long range
correlations) of what has been called nonextensive systems by the Tsallis
school \cite{Tsa01}.

\section{Application to biotechnology and biophysics}

In this last section, we give some examples of problems in the field of
biothechnology and biophysics, where we think the application of the
($n$,$\alpha)$ kinetics \ can open new paths to understand anomalous kinetics
from the point of view of the theory of complex systems.

\subsection{Sorption of dissolved contaminants in liquid phase}

The sorption (adsorption, chemisorption, biosorption) of pollutants from
aqueous solutions plays a significant role in water pollution control. It is
therefore important to be able to predict the rate at which contamination is
removed from aqueous solutions and how this rate depend on physical, chemical,
biological and environment variables in order to design an appropriate
treatment plant. Sorption of dissolved contaminant is a complex phenomena
caused by several mechanisms including London-van der Waals forces, Coulomb
forces, hydrogen bonding, ligand exchange fluctuations, chemisorption,
dipole-dipole forces and hydrophobic forces and biosorption for biological
materials. Therefore these systems can be considered to belong to the class of
complex systems \cite{Flo93}. The quantity adsorbed at time $t,$ $q_{t}$ is
defined as%
\begin{equation}
q_{t}=\frac{(c_{0}-c_{t})V}{W}%
\end{equation}
where $c_{0}$ is the initial concentration of the solution, $c_{t},$ the
concentration at time $t,$ $V,$ the volume of the solution and $W$, the weight
of the adsorbent.

In this context, the kinetic equations are determined with reference to the
quantity of dissolved contaminant necessary to reach equilibrium(for ex.
\cite{Ho03}). We have\ then
\begin{equation}
q_{t}=q_{e}(1-\exp(-K_{1}t)
\end{equation}
where $q_{e}\ $is the mass of solute adsorbed at equilibrium, $q_{t}$ is the
mass of solute adsorbed at time $t$ and $K_{1}$ is the rate coefficient.
Equation (43) is called pseudo first- order equation by contrast to the simple
exponential first-order equation (18).

In the same way, one can define a pseudo second-order reaction:%
\begin{equation}
\frac{1}{q_{e}-q_{t}}\ =K_{2}t+\frac{1}{q_{e}}%
\end{equation}
In agreement with the ideas developed in sections 1 and 2, we can introduce
the pseudo-($n$,$\alpha)$ equation%
\begin{equation}
q_{t}(\alpha,n)=q_{e}[1-(1+q_{e}^{n-1}(n-1)K_{n,\alpha}t^{\alpha})^{-\frac
{1}{n-1}}]
\end{equation}
which reduces to (43) for $n=1,\alpha=1$ and to (44) for $n=2,$ $\alpha=1$. We
can write (45) more compactly using the definition (4) of the deformed
exponential $\exp_{n}(x),$%
\begin{equation}
r_{n,\alpha}(t)=q_{t}(\alpha,n)/q_{e}=1-\exp_{n}((t/\tau_{n,\alpha})^{\alpha
})\text{ \ \ \ \ \ \ \ with \ \ \ \ \ }\tau_{n,\alpha}=(q_{e}^{n-1}%
K_{n,\alpha})^{-1/\alpha}%
\end{equation}
The definition of the deformed logarithm (4) associated with $\exp_{n}(x)$
allows us to write the following relation%
\begin{equation}
R(t)=\mathrm{Log}(\frac{(1-r_{n,\alpha}(t))^{1-n}-1}{n-1})=\alpha
\mathrm{Log}(t)-\alpha\mathrm{Log}(\tau_{n,\alpha})
\end{equation}
which can be used to make a linear fit of the data $(r_{n,\alpha}%
(t)=q_{t}(\alpha,n)/q_{e})$ and obtain the values of $\alpha$ and
$\tau_{n,\alpha}$. The value of $n$ to be chosen is the one which can give the
better fit in the Log-Log plot.\ 

In two different collaborations we have analyzed the kinetics of various
pollutants (phenol, tannic acid, gallic acid, melano\"{\i}dine on activated
carbon \cite{Gua05} and various dyes pollutants from the textile industry on
biological materials (algaes and agaves) \cite{Tun05}. It appears that, quite
generally the data can be fitted quite well to pseudo-($n$,$\alpha)$ kinetic.
For $t<<\tau$, the concentration $q_{e}(n,\alpha)$ does not depend on $n$ (cf
eq.16) and can be fitted to a $n$-independent power law $q_{e}(n,\alpha)$
$\varpropto t^{\alpha}$ .$\ $The value of $n$ (i.e. the overall order of the
reaction) has to be determined from the large time (near saturation) behavior
of the kinetics. Contrary to the usual practice in that field, we found that
the exponent $\alpha$, which is implicitly equal to 1 in the traditional
analysis of kinetic data \cite{Ho03} in terms of a pseudo-first or -second
order reaction, is more appropriate to characterize the kinetics of sorption
of dissolved contaminants in liquid phase. It yields a better fit and moreover
it is related to the system energy landscape which varies with external
parameters. In Fig.1, we show an example (adsorption kinetic on activated
carbon of m\'{e}lano\"{\i}dine, a dye formed in the crystallization process of
sacharose \cite{Gua05}) where the best fit is obtained with a pseudo-($1.5$%
,$\alpha)$ kinetic\ equation.\ One cannot fit properly the data with a
quasi-first or second-order kinetic where $\alpha=1$. This is understandable
since one observes that for small $t$ , $q_{e}(n,\alpha)$ $\varpropto
t^{0.56}$. The dependence of the two quantities $\alpha$ and $n$ on the
physical, chemical environmental and biological parameters of the couple
adsorbent-pollutant (pH, T, clustering, ligand field and architecture of large
biomolecules...) is the subject of current studies\ \cite{Gua05}\cite{Tun05}.

\subsection{Kinetics in photosynthesis processes}

In this subsection, we want to show how the fractal ($n,\alpha$) rate equation
(41) can be used in situation where a two-steps first order equation has been
used to fit the kinetics of photosynthesis processes. The example chosen is
the kinetics of the conversion of protochlorophyllide into chlorophyllide. The
method used to follows the kinetics in that problem is the observation of
spectral changes recorded by the technique of spectrofluorometry under
short-time illumination. For instance in \cite{Sir68} the authors have
observed the transformation of a 647 nm pigment (species a) by 630 nm
photons.\ Contrary to the transformation induced by 647 nm photons, where the
kinetics is first order, the transformation under 630 nm irradiation follows
an unusual kinetic. It has been assumed that this particular kinetics is due
to the presence of an other protochlorophyllide species (named b), i.e. a
pigment with another association with the lipoproteins. We refer to the
specialized literature for details \cite{Sir68}\cite{Bor62}\cite{Sir68}. The
two-steps model, often used when the rate equation cannot be fitted to a first
or second-order kinetics, lead Boardman, in this particular problem
\cite{Bor62}, to the following rate equation (percentage of phototransformed
quantity),%
\begin{equation}
T(\%)=100-A\exp(-K_{1}t)-(100-A)\exp(-K_{2}t)
\end{equation}
A is the proportion in \% of the complex which is transformed. This formula
can be fitted to the experimental data up to 85\% \cite{Sir68}. What differs
in 647 pigment and 630 nm pigment is the link with the lipoproteins. The
pigment-protein links are most probably fluctuating locally due to the
complexity of the organization of the molecules inside the prolamellar body.
If instead of two, a distribution of "species" is present, it is more
appropriate to use, a rate equation deriving from a distribution of
exponentials as the $(n,\alpha)$ rate equation. We have verified that in this
particular case, (42) can fit perfectly the experimental curve, also for
transformation larger than 85\% (Fig.1).\ The two asymptotic behaviors appears
to be power-law, a behavior a simple two-step mechanism cannot account for.
The best fit obtained with a nonlinear method is given by%
\begin{equation}
T(\%)=100(1-(1+(1.9-1)\ (t/0.07)^{0..96})^{-1/(1.9-1)}))
\end{equation}
This correspond to a fractal-time exponent $\alpha=0.96$ \ and a reaction
order parameter $n=1.9$. The characteristic time is $\tau_{\alpha}=0.07$ and
from the inspection of the experimental curve (Fig.1), one can see that it is
very close to $\tau_{1/2}.$ The linear fit ($r(t)=T/100$ ) gives the same
result (Fig.2):%
\begin{equation}
R(t)=\mathrm{Log}(\frac{(1-r(t))^{1-1.9}-1}{1.9-1})=0.96\mathrm{Log}%
(t)-0.96\mathrm{Log}(0.07)
\end{equation}
with a regression\ coefficient of 0.9997.

The temperature dependence of the transformation rate under 633 nm photons
\cite{Sir70} indicates that both endo- and exothermic effects are
competing.\ As we have suggested at the end of section 2, in this particular
situation, fluctuations of the rate coefficient in the exponent of the
Arrhenius law (29) can give rise to $(n,\alpha)$ kinetics.

\subsection{Complexity of DNA}

One method widely used to study the complexity of DNA is the so-called Cot
method.\ The method splits the double strands of DNA into single strands by
raising the temperature or by other denaturing process. One then studies the
kinetics of the reassociation of dissociated single strands$.$ Since it
involves two single strands, the renaturing into the original form is assumed
to follow a (2,1) kinetics $\ f=c\ (t)/c_{\ }(0)=[(1+\ c(0)(Kt)]^{-1} $.\ This
equation is the basis of the Cot analysis of the rate of renaturation of
sequence heterogeneity (or complexity) of DNA, The quantity $c(0)$ is the
initial concentration of DNA, \ $f$$=c\ (t)/c_{\ }(0)$ the fraction of
single-stranded molecules which decreases with time and $K$ the rate constant
for the reassociation of complementary strands. The value of $c(0)t$ \ when
$f\ =0.5$ is known as $c(0)t_{1/2}$.

The rate coefficient $K$ is characteristic of a particular DNA\ and is related
to its complexity in terms of sequence composition.\ The quantity
$c(0)t_{1/2}$ is the reciprocal of $K$ and can therefore be used as a measure
of sequence complexity.\ The higher the value of $c(0)t_{1/2},$ the more
complex is the DNA.

The Cot method which was developed in the 1960's and widely used in the 70'
was then nearly abandoned.\ It made a comeback recently \cite{Pat02} as a much
cheaper method because of its ability to concentrate on the low copy
sequences, the highly repeated sequences being irrelevant as far as the
genetic information is concerned.

Due to the complexity of the DNA structure, it would be surprising if there
would be no short time memory effects in the reassociation process, since it
involves quite complex biomolecules. Indeed, fingerprints of fractality and
nonextensivity in DNA fragment distribution has been reported recently
\cite{Sot04}. \ The ideas and formalism developed in this paper might be of
interest also in that field of primary importance.

\section{Discussion}

Using ideas and theoretical tools borrowed from recent works on the theory of
relaxation, we have derived a universal function for the kinetics of complex
systems characterized by stretched exponential and/or power-law behaviors This
kinetic function unifies and generalizes previous theoretical attempts to
describe what has been called "fractal kinetic". The concentration
evolutionary equation (12) is formally similar to the Burr$_{XII}$ relaxation
function obtained in the theory of relaxation, with two exponents $\alpha$ and
$n$.\ The first one is due to memory effects and short-range correlations and
the second one finds its origin in the long-range correlations and geometrical
frustrations which give rise to ageing behavior. As in the theory of
relaxation, these effects can be formally handled by introducing adequate
probability distributions for the rate coefficient.\ We have shown that the
distribution of rate coefficients is the consequence of local variations of
the free energy appearing in the exponent of the Arrhenius formula. The
scaling (power-law) behavior of the kinetic is therefore an other example of
what has been called "exponential conspiracy" \cite{Bou96}. The two
macroscopic observables $n$ and $\alpha$ are formally related to the energy
landscape of the complex system which varies if physical, chemical or
biological external factors are modified.

The fractal $(n,\alpha)$ kinetic has been applied to a few problems of
fundamental and practical importance \cite{Gua05}\cite{Tun05}\cite{Fra05},
examples of which have been presented in section 4.

In references \cite{Bro04} we have shown how a universal relaxation function
can be derived if we use distributions of macroscopic waiting times maximizing
the nonextensive Tsallis entropy.\ Similar conclusions can be drawn in the
present problem, if we introduce local reaction waiting time in a
probabilistic derivation of the universal kinetic function.

The relation between the $(n,\alpha)$ kinetic and nonextensive
thermostatistics will be the subject of further studies.

\section{Captions of Figures}

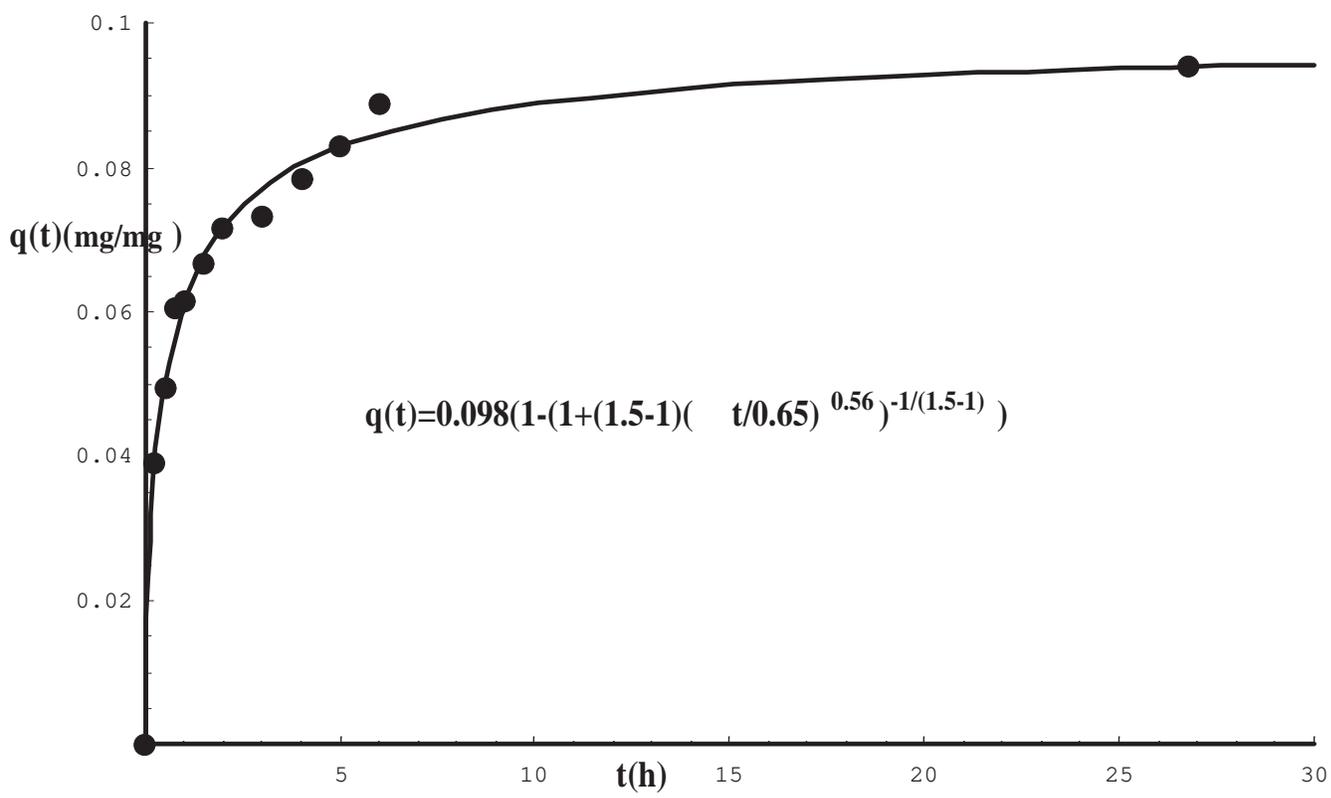
Fig.1 $(n,\alpha)$ kinetic of adsorption of melano\"{\i}dine in aqueous
solution on activated carbon with $n=1.5$ and $\alpha=0.56$.

Fig.2 $(n,\alpha)$ kinetic (eq.49) of the conversion of protochlorophyllide
into chlorophyllide (transformation of 647 nm pigment by 630 nm photons).\ The
smaller points for t%
$>$%
0.4 sec. are results of the two first order model (eq.48).

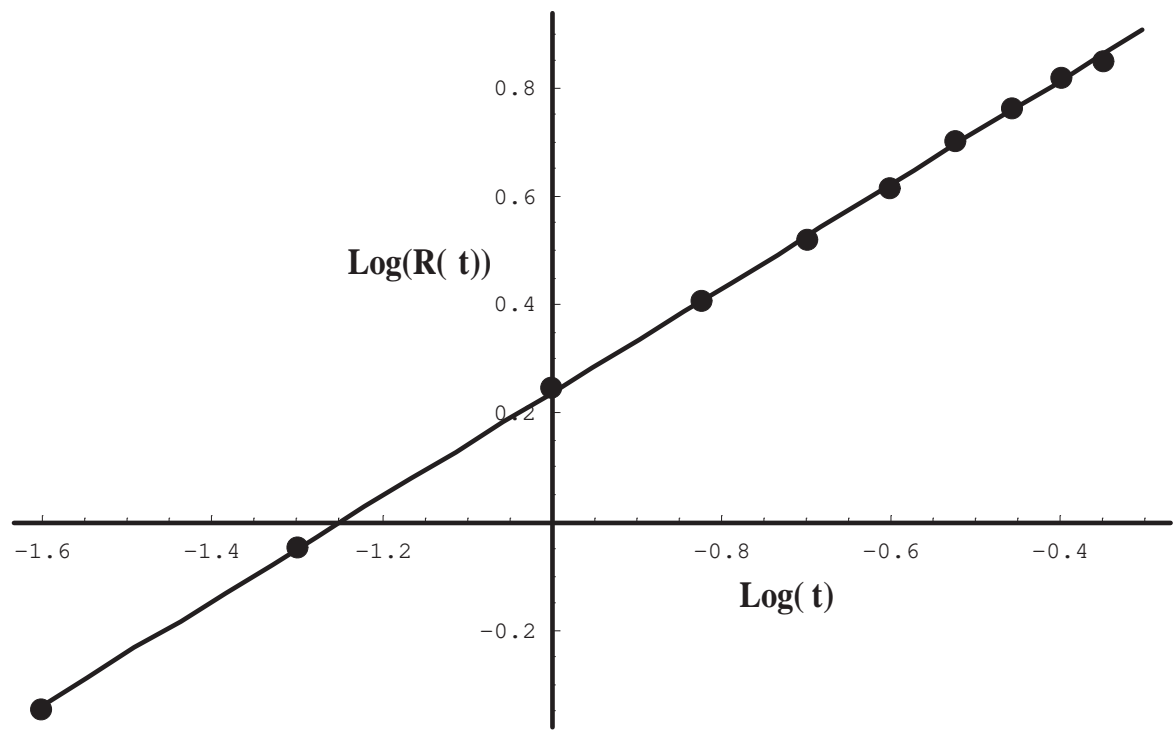
Fig.3 Log-Log plot (eq.47) applied to data of Fig.2.

\end{document}